\documentclass{emulateapj}
\usepackage{amssymb}
\usepackage{graphicx}

\newcommand{\be}{\begin{eqnarray}}
\newcommand{\ee}{\end{eqnarray}}

\begin{document}
\title{Radio Emission from Pulsar Wind Nebulae without Surrounding Supernova Ejecta: Application to FRB 121102}
\author{Z. G. Dai$^{1,2}$, J. S. Wang$^{1,2,3}$, \& Y. W. Yu$^{4,5}$}
\affil{$^1$School of Astronomy and Space Science, Nanjing University, Nanjing 210093, China; dzg@nju.edu.cn\\
$^2$Key Laboratory of Modern Astronomy and Astrophysics (Nanjing University), Ministry of Education, China\\
$^3$Max-Planck-Institut f\"ur Kernphysik, Saupfercheckweg 1, D-69117 Heidelberg, Germany\\
$^4$Institute of Astrophysics, Central China Normal University, Wuhan 430079, China\\
$^5$Key Laboratory of Quark and Lepton Physics (Central China Normal University), Ministry of Education, China}

\begin{abstract}
In this paper, we propose a new scenario in which a rapidly-rotating strongly-magnetized pulsar
without any surrounding supernova ejecta produces fast radio bursts (FRBs) repeatedly via some mechanisms, and meanwhile,
an ultra-relativistic electron/positron pair wind from the pulsar sweeps up its ambient dense
interstellar medium, giving rise to a non-relativistic pulsar wind nebula (PWN). We show that the synchrotron
radio emission from such a PWN is bright enough to account for the recently-discovered persistent
radio source associated with the repeating FRB 121102 in reasonable ranges of the model parameters. In addition,
our PWN scenario is consistent with the non-evolution of the dispersion measure inferred from
all the repeating bursts observed in four years.
\end{abstract}

\keywords{ pulsars: general -- radiation mechanisms: non-thermal -- radio continuum: general -- stars: neutron}

\section{Introduction}

Fast radio bursts (FRBs) are millisecond-duration flashes of coherent GHz radio emission of unknown physical origin \citep{Lorimer2007,Keane2012,Thornton2013,Spitler2014,Champion2015,Masui2015,Ravi2015,Ravi2016,Petroff2016,Spitler2016,Chatterjee2017}.
Most of them arise from high Galactic latitudes, but their inferred dispersion measures, {${\rm DM}\sim 300-1600\,{\rm pc\,cm^{-3}}$,
are much larger than expected for propagation through the cold plasma of our Galaxy and its halo, strongly suggesting that
they are at cosmological distances \citep[for a review on observations and models see][]{Katz2016a}.

Only one repeating case, FRB 121102, was first detected to occur on 2 November 2012 \citep{Spitler2014}. Surprisingly,
10, 6, and 13 additional bright bursts from the direction of this FRB were reported to appear
only in three different times, respectively \citep{Spitler2016,Scholz2016,Chatterjee2017,Marcote2017},
appearing to indicate the temporally-clustering feature of these repeating bursts. More importantly, the discovery of
both persistent radio and optical sources associated with FRB 121102 and the identification of a host dwarf galaxy
at a redshift of $z=0.193$ \citep{Chatterjee2017,Marcote2017,Tendulkar2017} certainly confirm a cosmological
origin of this FRB.

These observations rule out the catastrophic event models such as the collapse of supra-massive neutron stars
to black holes or the merger of binary compact objects. Four types of radio emission for FRB 121102 have been
discussed in detail. First, in the {\em rotationally-powered} model
\citep[e.g.,][]{Connor2016,Cordes2016,Lyutikov2016,Metzger2017,Kashiyama2017}, FRBs from a millisecond magnetar
are suggested to be a scaled-up version of super-giant pulses from the Crab pulsar.\footnote{\cite{Kisaka2017}
constrained the parameters of a pulsar powering FRB 121102 based on the giant-pulse emission model from
the luminosities and durations of the 30 observed bursts.} Second, in the {\em magnetically-powered} model
\citep[e.g.,][]{Popov2010,Kulkarni2014,Katz2016b,Metzger2017,Kashiyama2017}, FRBs may arise from the unexpected release
of magnetic energy \citep[or electrostatic energy see][]{Katz2017} in the magnetar's interior, similar to the giant flare model of
Galactic magnetars. FRBs might also occur repeatedly during the accretion of magnetized materials onto a neutron star
from its white dwarf companion \citep{Gu2016}. Third, in the {\em gravitationally-powered} model \citep{Dai2016}, repeating bursts
may originate from a strongly magnetized pulsar encountering an asteroid belt around another star. This model can
account for several previously-observed properties including the duration distribution, repetitive rate, and temporal clustering
of the bursts. Fourth, in the {\em kinetically-powered} ``cosmic-comb'' model \citep{Zhang2017}, FRBs may be produced in the magnetosphere
of a regular pulsar that is ``combed'' suddenly and repeatedly by a nearby strong plasma stream towards the anti-stream direction.
No matter which type of energy source of a FRB is correct, some stringent constraints on the spin period and surface magnetic field strength
of a central pulsar have been derived from the recent observations \citep[e.g.][]{Cao2017}.

While the physical origin of FRB 121102 remains controversial, the persistent radio emission source associated with this FRB,
which was recently discovered by \cite{Chatterjee2017} and further detected by \cite{Marcote2017}, becomes mysterious. \cite{Murase2016}
predicted the persistent radio emission from the termination shock produced by
the interaction of an ultra-relativistic pulsar wind with the supernova (SN) ejecta,\footnote{\cite{Yang2016} studied
the heating effect of an FRB on its ambient self-absorbed synchrotron nebula and found an obvious, detectable hump of
the nebula spectrum in several decades near the self-absorption frequency.} and very recently, \cite{Kashiyama2017} utilized
the observed radio data to constrain the parameters of this model. In addition, \cite{Metzger2017} explored the radio emission from
the forward shock produced by the interaction between the fast outer layer of SN ejecta with its ambient medium.
These works assumed the SN ejecta with a mass $\sim 10M_\odot$. It is so massive SN ejecta that would lead to
an observational evolution of DM over the year time scale for a very young age of a few decades \citep{Piro2016,Lyutikov2017,Metzger2017}.
However, the non-detection of DM evolution requires that the SN ejecta should have a much smaller mass. \cite{Kashiyama2017}
suggested one solution to this question, i.e., an ultra-stripped SN with a mass $\lesssim 0.1M_\odot$ is possibly associated
with FRB 121102. \cite{Piro2013} have studied the radio emission from the SN ejecta both that has such a small mass
and that is powered by a millisecond magnetar, and found an observational evolution of the radio emission flux over the year time scale.
It is unclear whether the persistent radio source associated with FRB 121102 shows a similar evolution.

In this paper, we propose a new scenario for the persistent radio source, in which a rapidly-rotating strongly magnetized pulsar
is not surrounded by the SN ejecta. Such a situation may appear if a pulsar has an extremely high kick velocity
to leave away far from its birth site \citep{Chatterjee2004,Hobbs2005} or if a pulsar escapes from its high-mass X-ray binary system
during the explosion of its companion star \citep{Bhat1991} or if a pulsar is born (and then moves away) during the merger of
binary neutron stars \citep{Dai2006,Gia2013,Yu2015} or the accretion-induced collapse of a white dwarf \citep{Canal1976,Nomoto1991,Yu2015}.
While this pulsar may produce bursts repeatedly through some mechanisms mentioned above, an ultra-relativistic wind from the pulsar
is sweeping up its ambient dense interstellar medium, giving rise to a non-relativistic pulsar wind nebula
(hereafter PWN) {\em without surrounding SN ejecta}. We show that our PWN scenario
can explain the persistent radio source in reasonable ranges of the model parameters. This paper is organized as follows.
In Section 2, we analyze the dynamics of the PWN, and in Section 3, we discuss the properties of synchrotron radio emission from the PWN.
In Section 4, we constrain the model parameters and discuss the DM contributed by the PWN and innermost cold wind. Finally,
in Section 5, we present our conclusions and discussion.

\section{Dynamics of a PWN without Surrounding Ejecta}

A highly-magnetized pulsar generates a cold ultra-relativistic wind dominated by electron/positron pairs (maybe
including a very small number of baryons) with a luminosity of $L_w$ and a bulk Lorentz factor of $\Gamma_w$. This wind sweeps up
an ambient dense medium, leading to two shocks: a reverse shock (i.e., a termination shock with a radius of $R_t$)
that propagates into the cold wind and a forward shock that propagates into the ambient medium.
Thus, the system has a four-zone structure consisting of (1) outermost, an unshocked
medium with a constant number density of $n_0$, (2) next, a forward-shocked medium,
(3) a reverse-shocked wind gas (i.e., a PWN without surrounding SN ejecta), and (4) innermost,
an unshocked cold wind from the pulsar, where regions 2 and 3 are separated by a contact discontinuity with
a radius of $R_p$. By assuming that a gamma-ray burst is driven by a newborn millisecond magnetar, \cite{Dai2004}
studied observational signatures of a post-burst relativistic PWN powered by such a magnetar,
and found a plateau in the light curve of an early afterglow due to the reverse shock emission. This feature
provides an explanation for the light-curve plateaus of gamma-ray burst afterglows observed by Swift \citep{Yu2007}.
To explain non-repeating FRBs, \cite{Lyubarsky2014} discussed a PWN (without surrounding SN ejecta) powered
by a slow-rotating magnetar with a typical period of a few seconds, and suggested that the interaction of
a giant-flare magnetic pulse from the magnetar with such a PWN could lead to a FRB via synchrotron maser emission
from relativistic shocks. Although this model of \cite{Lyubarsky2014} cannot account for FRB 121102-like
repeating bursts due to an extremely low rate ($\sim$ one per magnetar per four decades) of the observed giant flare
events in our Galaxy, the model predicts a persistent synchrotron radio emission from the PWN. The flux density
of such an emission has an upper limit of $\sim L_{\rm sd}/(4\pi D_L^2\nu)\sim 0.03[L_{\rm sd}/(4\times 10^{34}
\,{\rm erg}\,{\rm s}^{-1})](D_L/1\,{\rm Gpc})^{-2}(\nu/1\,{\rm GHz})^{-1}\,\mu{\rm Jy}$, where $L_{\rm sd}$ is
the spin-down luminosity of the slow-rotating magnetar. Therefore, this emission would be undetectable
for the PWN at a cosmological distance.  In this paper, we investigate the persistent synchrotron
radio emission from a non-relativistic PWN powered by a rapidly-rotating highly-magnetized pulsar and show that
this emission would be observable even if the PWN is at a cosmological distance.

We first discuss how the system evolves dynamically with time. On one hand, while the heating mechanism of the PWN (region 3)
is continuous energy injection from the pulsar, the dominant energy loss of the PWN is work against the forward-shocked medium (region 2),
so that the total energy $E_3$ of the PWN evolves through
\begin{equation}
\frac{dE_3}{dt}=L_w-4\pi R_p^2P_2\frac{dR_p}{dt}\label{E3t},
\end{equation}
and
\begin{equation}
E_3=\left(\frac{4\pi}{3} R_p^3\right)\times (3P_3)\label{E3},
\end{equation}
where $t$ is the dynamically-expanding time of the PWN, $P_2$ and $P_3$ are the pressures of regions 2 and 3 respectively, and $P_2=P_3$
on both sides of the contact discontinuity. Please note that the first factor (volume) on the right side term of equation (\ref{E3})
is taken by assuming $R_t\ll R_p$ and that the second factor is the total energy density of the PWN, $U_3=3P_3$.

On the other hand, owing to the work from region 3 and the thin-shell approximation of region 2, the motion of region 2 follows from
\begin{equation}
\frac{d}{dt}\left[M_{sw}\frac{dR_p}{dt}\right]=4\pi R_p^2P_3\label{Mswt},
\end{equation}
and
\begin{equation}
M_{sw}=\frac{4\pi}{3}R_p^3n_0m_p\label{Msw}
\end{equation}
is the swept-up medium mass, where $m_p$ is the proton mass.
Thus, a combination of equations (\ref{E3t}) to (\ref{Msw}) gives
\begin{equation}
\frac{d}{dt}\left[R_p^2\frac{d}{dt}\left(M_{sw}\frac{dR_p}{dt}\right)\right]=R_pL_w\label{Rpt}.
\end{equation}

Assuming that $L_w$ is constant during the pulsar's spin-down timescale $t_{\rm sd}$, we obtain a solution to equation (\ref{Rpt}),
\begin{equation}
R_p=C\left(\frac{L_wt^3}{n_0m_p}\right)^{1/5}
=1.3\times 10^{18}\left(\frac{L_{w,41}t_2^3}{n_{0,2}}\right)^{1/5}{\rm cm}\label{Rp},
\end{equation}
where $C\equiv [125/(224\pi)]^{1/5}=0.708$, $L_{w,41}=L_w/10^{41}{\rm erg}\,{\rm s}^{-1}$, $t_2=t/10^2{\rm yr}$,
and $n_{0,2}=n_0/10^2{\rm cm}^{-3}$. This dynamics is similar to that of interstellar wind bubbles \citep{Castor1975}.
From equations (\ref{E3}), (\ref{Mswt}) and (\ref{Rp}), therefore, we can calculate the total energy and energy density of the PWN,
\begin{equation}
E_3=\frac{28\pi}{25}C^5L_wt=2.0\times 10^{50}L_{w,41}t_2\,{\rm erg}\label{totE},
\end{equation}
and
\begin{equation}
U_3=\frac{E_3}{(4\pi/3)R_p^3}=2.3\times 10^{-5}L_{w,41}^{2/5}n_{0,2}^{3/5}t_2^{-4/5}\,{\rm erg}\,{\rm cm}^{-3}\label{EDen}.
\end{equation}
According to \cite{Gaensler2006}, we obtain the radius of the termination shock
\begin{equation}
R_t\simeq \left(\frac{L_w}{4\pi cP_3}\right)^{1/2}=1.9\times 10^{17}L_{w,41}^{3/10}n_{0,2}^{-3/10}t_2^{2/5}\,{\rm cm}\label{Rt},
\end{equation}
where $c$ is the speed of light. It can be seen from equations (\ref{Rp}) and (\ref{Rt}) that the assumption $R_t\ll R_p$
is indeed valid if typical values of the model parameters are taken.

\section{Synchrotron Radio Radiation from the PWN}

We next discuss synchrotron radio radiation from the PWN. Electrons (and positrons) in the cold pulsar wind
(region 4) are accelerated to ultra-relativistic energies by the termination shock at $R_t$ and fill the PWN out to $R_p$.
We assume that their power-law spectrum behind the shock front is $dn_e/d\gamma_e=K\gamma_e^{-p}$ in units of electrons\,cm$^{-3}$.
Their synchrotron emission spectrum depends on three break frequencies. We consider the hard electron spectrum (i.e., $1<p<2$) in this paper.

The first break frequency is the synchrotron cooling frequency at which an electron with the cooling Lorentz factor $\gamma_c$
loses its energy in a dynamical time $t$. From \cite{Sari1998}, we get the cooling Lorentz factor
\begin{equation}
\gamma_c=\frac{6\pi m_ec}{\sigma_TB^2t}=4.3\times10^2\epsilon_B^{-1}L_{w,41}^{-2/5}n_{0,2}^{-3/5}t_2^{-1/5}\label{gammac},
\end{equation}
where $m_e$ is the electron mass, $\sigma_T$ is the Thomson cross-section, and $B=(8\pi \epsilon_BU_3)^{1/2}
=2.4\times10^{-2}\epsilon_B^{1/2}L_{w,41}^{1/5}n_{0,2}^{3/10}t_{2}^{-2/5}$\,G is
the magnetic field strength in the PWN under the assumption that the magnetic energy density behind the shock
is a fraction $\epsilon_B$ of the total energy density. Thus, the synchrotron cooling frequency is calculated by
\begin{equation}
\nu_c=\gamma_c^2\frac{q_eB}{2\pi m_ec}=1.2\times 10^{10} \epsilon_B^{-3/2}L_{w,41}^{-3/5} n_{0,2}^{-9/10} t_2^{-4/5}\,{\rm Hz}\label{nuc},
\end{equation}
where $q_e$ is the electron charge.

Owing to this cooling effect, the electron spectrum behind the termination shock becomes \citep{Sari1998}
\begin{equation}
\frac{dn_e}{d\gamma_e}=\left \{
       \begin{array}{ll}
         K\gamma_e^{-p}, & \gamma_{\rm min}\le\gamma_e<\gamma_c,\\
         K\gamma_c\gamma_e^{-(p+1)}, & \gamma_c\le\gamma_e\le \gamma_{\rm max},
        \end{array}
       \right.\label{dne}
\end{equation}
where $\gamma_{\rm min}$ and $\gamma_{\rm max}$ are the minimum and maximum Lorentz factors of the shock-accelerated electrons, respectively.
Here we only discuss the slow-cooling regime to account for the spectrum of the persistent radio source associated with FRB 121102.
In the following calculations, we fix $p=1.4$ \citep{Chatterjee2017}.

We further assume that the electron energy density behind the shock is a fraction $\epsilon_e$ of the total energy density,
\begin{equation}
U_e=\epsilon_eU_3=\int_{\gamma_{\rm min}}^{\gamma_{\rm max}}\left(\frac{dn_e}{d\gamma_e}\right)(\gamma_e m_ec^2)d\gamma_e\label{Ue}.
\end{equation}
Please note that $\epsilon_e+\epsilon_B=1$ in our PWN scenario. Inserting equation (\ref{dne}) into equation (\ref{Ue}), we find
\begin{eqnarray}
K & = & \frac{(2-p)(p-1)\epsilon_eU_3}{m_ec^2\gamma_c^{2-p}}\nonumber \\
& = & 0.18\epsilon_e\epsilon_B^{3/5}L_{w,41}^{16/25}n_{0,2}^{24/25}t_2^{-17/25}\,{\rm cm}^{-3}\label{Kvalue}.
\end{eqnarray}
The second break frequency is the typical synchrotron frequency which an electron with $\gamma_{\rm min}$ radiates,
\begin{eqnarray}
\nu_m & = & \gamma_{\rm min}^2\frac{q_eB}{2\pi m_ec}\nonumber \\
& = & 6.7\times10^4\epsilon_B^{1/2}\gamma_{\rm min}^2L_{w,41}^{1/5}n_{0,2}^{3/10} t_2^{-2/5}\,{\rm Hz}\label{num}.
\end{eqnarray}

The third break frequency is the synchrotron self-absorption frequency \citep{Wu2003},
\begin{eqnarray}
\nu_a & = & \left(\frac{c_2q_eKR_p}{B}\right)^{2/(p+4)}\frac{q_eB}{2\pi m_ec}\nonumber \\
& = & 6.8\times10^8\epsilon_e^{10/27}\epsilon_B^{29/54}\nonumber \\ & & \times L_{w,41}^{59/135}n_{0,2}^{127/270}t_2^{-38/135}\,{\rm Hz}\label{nua2}
\end{eqnarray}
for $\nu_m<\nu_a<\nu_c$, where the coefficient $c_2$ depends on $p$ \citep[see Appendix A of][]{Wu2003}.

The peak flux density at a luminosity distance of $D_L$ from the source is calculated by \citep{Sari1998}
\begin{eqnarray}
F_{\nu,\rm max} & = & \frac{N_e}{4\pi D_L^2}\frac{m_ec^2\sigma_T}{3q_e}B\nonumber \\
& = & 3.1\times 10^4\epsilon_e\epsilon_B^{11/10}\gamma_{\rm min}^{-(p-1)}\nonumber \\ & & \times
L_{w,41}^{36/25}n_{0,2}^{33/50}t_2^{18/25}\,\mu{\rm Jy}\label{Fmax},
\end{eqnarray}
where $N_e=4\pi R_p^3K/[3(p-1)\gamma_{\rm min}^{p-1}]$ is the total electron number of the PWN.
The synchrotron emission flux density at any frequency $\nu$ is given by \citep{Meszaros1997,Sari1998}
\begin{equation}
F_\nu= \left \{
       \begin{array}{lll}
       F_{\nu,\rm max}(\nu_a/\nu_m)^{-(p-1)/2}(\nu/\nu_a)^{5/2}, & \nu<\nu_a,\\
       F_{\nu,\rm max}(\nu/\nu_m)^{-(p-1)/2}, & \nu_a<\nu<\nu_c,\\
       F_{\nu,\rm max}(\nu_c/\nu_m)^{-(p-1)/2}(\nu/\nu_c)^{-p/2}, & \nu\ge\nu_c.
       \end{array}
       \right.\label{Fnu}
\end{equation}
After inserting equations (\ref{num}) and (\ref{Fmax}) into (\ref{Fnu}), {\em it is interesting to note that $F_\nu$
is independent of $\gamma_{\rm min}$ for any value of $p$.} Thus, we can compare our PWN scenario with the observations
on the persistent radio source associated with FRB 121102 to constrain four remaining parameters ($L_w$, $n_0$,
$\epsilon_B$, and $t$) in the next section.

\section{Constraints on the Model Parameters}

The Very Large Array-observed spectrum ($F_\nu$) of the persistent radio source associated with FRB 121102
\citep[see Extended Data Figure 2 of][]{Chatterjee2017}
indicates the spectral index $\alpha\sim -0.2$ for $\nu\lesssim 10\,$GHz and $\alpha\sim -0.8$ for $\nu\gtrsim 10\,$GHz.
Compared with equation (\ref{Fnu}), this emission spectrum is consistent with the hard electron spectrum $p\sim 1.4$, and
thus the observations require that (i) $\nu_c\simeq 10\,$GHz, (ii) $F_{\rm 10\,GHz}\simeq 200\,\mu$Jy, and (iii) $\nu_a\lesssim 1.4\,$GHz,
in our PWN scenario.

The other requirements are as follows: (iv) The size of the PWN should be smaller than the observed upper limit on the size of
the persistent radio source \citep{Marcote2017}, $R_p\lesssim 0.7\,$pc. (v) The radius of the termination shock, $R_t$, must
be much smaller than the radius of the contact discontinuity, $R_p$, in order that our PWN scenario is self-consistent.
(vi) The DM contributed from the shocked medium should be smaller than the estimated host-galaxy DM
\citep{Tendulkar2017,Cao2017,Yang2017}, ${\rm DM}_{\rm ISM}=n_0R_p\lesssim{\rm DM}_{\rm host}\sim 100\,$pc\,cm$^{-3}$.
(vii) The age of the PWN should be larger than the total observation period of time, $t\gtrsim4\,$yr.

\begin{figure}
\begin{center}
\includegraphics[scale=0.51]{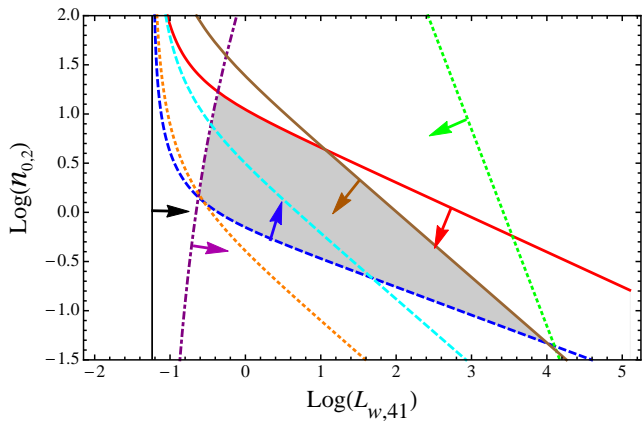}
\caption{Constraints on $L_{w,41}$ and $n_{0,2}$ from requirements (iii, red solid line), (iv, blue dashed line),
(v, green dotted line), and (vi, purple dot-dashed line), and (vii, brown solid line), as shown in the text.
The cyan dashed line and orange dotted line are plotted based on equation (\ref{t2})
for $t=40$ and $400\,$yr, respectively. The black solid line is corresponding to $\epsilon_B>0$.
The shaded region includes the permitted values of $L_{w,41}$ and $n_{0,2}$.}
\end{center}
\end{figure}

According to the seven requirements listed above, we can constrain $L_{w,41}$ and $n_{0,2}$. Figure 1 presents these constraints
on the $L_{w,41}$ and $n_{0,2}$ plane. On one hand, from requirements (i) and (ii), we obtain
\begin{equation}
\epsilon_B\simeq 1-0.06L_{w,41}^{-1}\label{eB},
\end{equation}
where $L_{w,41}>0.06$ must be satisfied (as shown in Figure 1) so that $\epsilon_B>0$, and
\begin{equation}
t_2\simeq 1.3L_{w,41}^{-3/4}n_{0,2}^{-9/8}\epsilon_B^{-15/8}\label{t2}.
\end{equation}
On the other hand, by considering requirements (iii)-(vii), we obtain the constraints on $L_{w,41}$ and $n_{0,2}$
from requirements (iii, red solid line), (iv, blue dashed line), (vi, purple dot-dashed line), and (vii, brown solid line).
The shaded region in Figure 1 includes the permitted values of $L_{w,41}$ and $n_{0,2}$. In addition, once $L_{w,41}$ and $n_{0,2}$ are given,
$\epsilon_B$ and $t$ can be calculated from equations (\ref{eB}) and (\ref{t2}).

Now let's further discuss constraints on the period and surface dipole magnetic field strength of the pulsar
for given $L_{w,41}$ and $t_2$. We assume that $P_*$ is the initial period of the pulsar when it starts to drive the PWN, $I$ is
its moment of inertia, $B_*$ is the pulsar's surface dipole magnetic field strength, and $R_*$ is the stellar radius.
The pulsar's spin-down luminosity and timescale due to magnetic dipole radiation are estimated by
\begin{equation}
L_{\rm sd}= 3.8\times 10^{43}B_{*,12}^2P_{*,-3}^{-4}R_{*,6}^6\,{\rm erg}\,{\rm s}^{-1}\label{Lsd},
\end{equation}
and
\begin{equation}
t_{\rm sd}= 16B_{*,12}^{-2}P_{*,-3}^2I_{45}R_{*,6}^{-6}\,{\rm yr}\label{tsd},
\end{equation}
respectively, where $B_{*,12}=B_*/10^{12}\,{\rm G}$, $P_{*,-3}=P_*/1\,$ms, $I_{45}=I/10^{45}\,{\rm g}\,{\rm cm}^2$,
and $R_{*,6}=R_*/10^6\,{\rm cm}$. If $t\lesssim t_{\rm sd}$ and $L_w=L_{\rm sd}\simeq {\rm constant}$ are required
to guarantee the validity of equation (\ref{Rp}),
then we find
\begin{equation}
P_{*,-3}\lesssim 7.8L_{w,41}^{-1/2}t_2^{-1/2}I_{45}^{1/2}\label{Period},
\end{equation}
and
\begin{equation}
B_{*,12}\lesssim 3.2L_{w,41}^{-1/2}t_2^{-1}I_{45}R_{*,6}^{-3}\label{Field}.
\end{equation}
This constraint on $B_*$ for $L_w\gtrsim10^{41}\,{\rm erg}\,{\rm s}^{-1}$ is not inconsistent with the limits based on the rotationally-powered model
\citep[see equation 7 in][]{Lyutikov2017} and the gravitationally-powered model \citep[see inequalities 9 and 16 in][]{Dai2016} of FRBs.
Of course, there is no limit on $B_*$ in the magnetically-powered model, provided that the average magnetic field strength
in the pulsar's interior is high enough \citep[e.g.][]{Metzger2017}.

In the above calculations, we have not taken into account any contribution of the pulsar wind regions (including the PWN
and innermost cold wind) to the DM of FRB 121102. In fact, a large number of electrons and positrons are required
in the PWN to produce the radio emission. The density of these leptons can be estimated to be $n_e=(2m_p/m_e)n_0/\Gamma_w$
by considering the pressure balance on two sides of the contact discontinuity,
$P_3\equiv(1/3)\times 4\Gamma_wn_em_ec^2=P_2\equiv(2/3)\times 4n_0m_pc^2$.
As a result, the DM contributed by the PWN is about ${\rm DM_{PWN}}=n_eR_p=15L_{w,41}^{1/5}n_{0,2}^{4/5}t_{2}^{3/5}
\Gamma_{w,4}^{-1}\,$pc\,cm$^{-3}$, which is basically consistent with the upper limit
of ${\rm DM_{PWN,max}}\lesssim100\,$pc\,cm$^{-3}$, where $\Gamma_{w,4}=\Gamma_w/10^4$. However, as pointed out by \cite{Cao2017},
a large number of leptons should come from a much smaller radius ($\lesssim R_{t}$) and even from the light cylinder of the pulsar,
where the lepton density and the Lorentz factor are much higher and thus a higher DM could be caused.
In particular, from \cite{Yu2014} and  \cite{Cao2017}, a stringent constraint on the spin period can be found
by requiring the DM of the total free wind to be smaller than the upper limit of ${\rm DM_{w, max}}$, that is,
$P_{*,-3}\gtrsim 6.0\mu_{\pm, \rm lc}^{2/3}L_{w,41}^{2/3}R_{*,6}^{-4}{\rm DM^{-1}_{w,max,2}}$, where $\mu_{\pm, \rm lc}$
is the multiplicity that represents the ratio of the wind lepton flux at the light cylinder to the Goldreich-Julian flux,
and ${\rm DM_{w,max,2}}={\rm DM_{w,max}}/10^2\,{\rm pc}\,{\rm cm}^{-3}$. This constraint on $P_*$
for $L_w\gtrsim10^{41}\,{\rm erg}\,{\rm s}^{-1}$ is basically in agreement with inequality (\ref{Period}),
if the DM contribution of the pulsar wind is comparable to that of the host galaxy and if the lepton density at the light cylinder
does not significantly deviate from the Goldreich-Julian density.

\section{Conclusions and Discussion}

In this paper we have proposed a new scenario for the recently-discovered persistent radio source associated with FRB 121102,
in which a rapidly-rotating strongly-magnetized pulsar has not been surrounded by the SN ejecta. This pulsar may produce
bursts repeatedly through the rotationally-powered or magnetically-powered or gravitationally-powered mechanisms listed
in the introductional section, and meanwhile, an ultra-relativistic electron/positron
pair wind from the pulsar interacts with its ambient dense interstellar medium, leading to a non-relativistic PWN without
surrounding SN ejecta. We studied the dynamics and synchrotron radio emission from such a PWN in detail. By fitting the observed
radio spectrum, we constrained the model parameters and found that all the parameters are in their reasonable ranges.
Therefore, our PWN scenario can provide an explanation for the persistent radio source associated with FRB 121102.
Furthermore, from requirement (vi) and discussion in Section 4, the time derivative of the DM contributed from the source,
$d{\rm DM_{src}}/dt\lesssim {\rm DM_{src, max}}/t\sim 1{\rm DM_{src, max,2}}t_2^{-1}\,{\rm pc}\,{\rm cm}^{-3}\,{\rm yr}^{-1}$
(where ${\rm DM_{src, max}}$ is the maximum DM from the source, including the contributions of the innermost free wind, PWN,
and shocked medium). This rate of DM change is undetectable and thus consistent with the non-evolution of the DM
inferred from all the repeating bursts observed in four years.

Finally, we give an order-of-magnitude estimate of the occurrence rate of persistent radio sources
from PWNe driven in dense interstellar environments. It is seen from inequality (\ref{Period}) that the period of a pulsar
$P_*\lesssim 10\,$ms to produce an observable cosmological PWN. In addition, the constraint on $B_*$ from inequality (\ref{Field}) is satisfied
for typical isolated pulsars and thus is not considered in the following discussion. The ratio of the number of such rapidly-rotating
pulsars to the total number of isolated pulsars in our Galaxy is estimated by $\xi(y)\sim \int_0^yf(x)dx/\int_0^\infty f(x)dx$,
where the period distribution function of isolated pulsars $f(x)\propto x^{a-1}e^{-x}$ and $x=P_*/P_0$ with two fitting parameters of
$a$ and $P_0$ \citep{Gil1996,Zhang2003}. If $y=0.01\,{\rm s}/P_0$ is taken, then we find that $\xi$ is in the range of
$\sim 6\times 10^{-5}$ to $\sim 2\times 10^{-4}$ by adopting different values of $a$ and $P_0$ from \cite{Zhang2003}
and \cite{Gil1996}. If this range of $\xi$ is reasonable for the other galaxies, therefore, the occurrence rate
of PWNe powered by rapidly-rotating strongly-magnetized pulsars can be approximately calculated by
${\cal R}_{\rm PWN}\sim \xi N_*N_{\rm gal}/t_{\rm H}\sim 10^3(\xi/10^{-4})(N_*/10^8)(N_{\rm gal}/10^9)\,{\rm yr}^{-1}$,
where $t_{\rm H}$ is the Hubble timescale, $N_*$ is the number of isolated pulsars in a galaxy, and $N_{\rm gal}$ is the number
of late-type galaxies within the cosmological comoving volume at redshift $z\lesssim 1$
\citep[for $N_*$ and $N_{\rm gal}$ also see Table 1 in][]{Dai2016}. It is interesting to note that
this rate has the same order of magnitude as the occurrence rate of repeating FRB sources estimated by 
\cite{Dai2016} within the frame of the pulsar-asteroid belt impact model.

\acknowledgements
We thank Bing Zhang for helpful comments and suggestions. This work was supported
by the National Basic Research Program (``973'' Program) of China (grant No. 2014CB845800)
and the National Natural Science Foundation of China (grant Nos. 11473008 and 11573014).

\end{document}